\documentstyle[12pt]{article}

\catcode`@=11
\def\citer{\@ifnextchar [{\@tempswatrue\@citexr}{\@tempswafalse\@citexr[]}}

\def\@citexr[#1]#2{\if@filesw\immediate\write\@auxout{\string\citation{#2}}\fi
  \def\@citea{}\@cite{\@for\@citeb:=#2\do
    {\@citea\def\@citea{--\penalty\@m}\@ifundefined
       {b@\@citeb}{{\bf ?}\@warning
       {Citation `\@citeb' on page \thepage \space undefined}}%
\hbox{\csname b@\@citeb\endcsname}}}{#1}}
\catcode`@=12

\newcommand{\lsim}{\raisebox{-0.13cm}{~\shortstack{$<$ \\[-0.07cm] $\sim$}}~}

\begin{document}

\begin{titlepage}

\begin{flushleft}
DESY 95--041\hfill ISSN~0418--9833 \\
FERMILAB--PUB--95/081--T \\
MPI/PhT/95--21 \\
hep-ph/9505225 \\
May 1995 \\
\end{flushleft}

\vspace{1.5cm}

\begin{center}

{\Large\sc Low--Energy Theorems in Higgs Physics}

\vspace{2cm}

{\large Bernd A. Kniehl$^{1,\,}$\footnote[2]{Permanent address:
Max--Planck--Institut f\"ur Physik, Werner--Heisenberg--Institut,
F\"ohringer Ring~6, 80805 Munich, Germany.}
and Michael Spira$^3$}

\vspace{1.5cm}

$^1$ Theoretical Physics Department, Fermi National Accelerator Laboratory,\\
P.O. Box 500, Batavia, IL 60510, USA\\
$^3$ II.~Institut f\"ur Theoretische Physik,\footnote[4]{Supported by
Bundesministerium
f\"ur Bildung und Forschung (BMBF), Bonn, Germany, under Contract 05
6 HH 93P (5) and by EU Program {\it Human Capital and Mobility} through
Network {\it Physics at High Energy Colliders} under Contract CHRX--CT93--0357
(DG12 COMA).} Universit\"at Hamburg,\\ Luruper Chaussee 149,
22761 Hamburg, Germany

\end{center}

\vspace{2cm}

\begin{abstract}
\noindent
We present low--energy theorems for the calculation of loop amplitudes with
external scalar or pseudoscalar Higgs bosons which are light compared to the
loop particles.
Starting from existing lowest--order versions of these theorems, we show how
their applicability may be extended to the two--loop level.
To illustrate the usefulness of these theorems, we discuss a number of
applications to Higgs production and decay at and beyond the one--loop order.
\end{abstract}

\end{titlepage}

\section{Introduction}
%        ============
The search for the scalar Higgs boson of the Standard Model
(${\cal SM}$) is one of
the most important tasks to be performed at present and future high--energy
experiments.  The only unknown independent parameter of this particle is
its mass, $M_H$.
{}From the direct search with the CERN Large Electron Positron Collider
(LEP1) and the SLAC Linear Collider (SLC) via the process $e^+e^-\to Z
\to Z^* H$, a lower limit on $M_H$ of 63.9~Ge$\!$V~has been obtained
at the 95\% confidence level \cite{janot}.
There are general
theoretical restrictions on the possible range of $M_H$.
Unitarity arguments lead to an upper bound of $\sim 700$~Ge$\!$V, if the
${\cal SM}$
is weakly interacting up to scale $\sim 1$~Te$\!$V;
this value comes down to $\sim 200$~Ge$\!$V, if
the ${\cal SM}$
is assumed to be valid up to the GUT
scale $\sim 10^{15}$~Ge$\!$V~\cite{unit}. On the other hand, the
requirement that the
${\cal SM}$ vacuum be stable sets a lower bound on $M_H$.
Assuming the ${\cal SM}$
to be valid up to scale $\Lambda \sim 1$~Te$\!$V and using
$m_t = (176 \pm 13)$~Ge$\!$V \cite{cdf} for the top--quark mass,
this lower bound amounts to about 55~Ge$\!$V, whereas for
$\Lambda \sim 10^{15}$~Ge$\!$V it is shifted to $\sim 130$~Ge$\!$V
\cite{vacstab}.
Recently, it has been pointed out that this $M_H$ lower bound is significantly
decreased by taking into account the possibility that the physical minimum of
the effective $\cal SM$ potential is metastable \cite{espi}.

It is attractive to study the minimal supersymmetric extension of the
${\cal SM}$ (${\cal MSSM}$).
It predicts five physical Higgs bosons:
two neutral ($\cal CP$--even) scalars ($h$ and $H$), one neutral
($\cal CP$--odd) pseudoscalar ($A$), and two charged scalars ($H^\pm$).
The mass of the lightest scalar ($h$) is restricted
to be below $\sim 140$~Ge$\!$V \cite{mh140}, whereas those of the heavy
scalars and the pseudoscalar will be typically of the order of the Fermi
scale, $v=246$~Ge$\!$V. The direct
search at LEP1 has excluded scalar--Higgs--boson masses below
$\sim 45$~Ge$\!$V and pseudoscalar--Higgs--boson masses below
$\sim 25$~Ge$\!$V \cite{mssmlep}.

If the Higgs bosons are lighter than the top quark and the $Z$ and $W$ bosons,
the latter may be integrated out.
In this way, the original Lagrangians describing the interactions of the
Higgs bosons with these heavy particles get replaced by effective Lagrangians.
These effective Lagrangians provide useful approximations for the interactions
of Higgs bosons in the low and intermediate mass range, below $\sim2M_Z$,
where at least one of the ${\cal SM}$ or ${\cal MSSM}$ Higgs bosons should
be found. The derivation of these
effective Lagrangians can be simplified by using low--energy
theorems (LETs) appropriate to external Higgs bosons with vanishing momentum.
This is the topic of the present article.

This paper is organized as follows. In Section~2, the LETs
for scalar and pseudoscalar Higgs bosons that are lighter than
the loop particles will be formulated at the lowest order of the perturbative
expansion.
These will then be generalized to higher orders, appropriate to
the application to multi--loop contributions.
In Section~3, we shall present applications of the theorems
to Higgs--boson production and decay processes within the ${\cal SM}$ and the
${\cal MSSM}$ at the one--loop level.
These examples will then be extended in Section~4 so as to include two--loop
corrections.
Section~5 will summarize our main results.

\section{Low--Energy Theorems}
%        ====================
In this section, LETs for any type of neutral
Higgs boson, generically denoted $\phi$, will be derived in the
limit of vanishing four--momentum $p_\phi$.
In this case, the Higgs boson acts as a constant field,
since $[{\cal P}_\mu,\phi] = i\partial_\mu \phi = 0$, with ${\cal P_\mu}$
being the four--momentum operator. As a consequence, the kinetic terms of
the Higgs Lagrangian vanish in this limit.

\subsection{Scalar Higgs Bosons}
%           ===================
In the $\cal SM$ and $\cal MSSM$, the Lagrangian for the interaction of
the neutral scalar Higgs boson(s) with the massive fermions and intermediate
bosons, having masses $m_i~(i=f,V)$, may be generated by the substitution
\citer{hsubst1,theorem2}
\begin{equation}
m_i \to m_i \left( 1 + \sum_\phi g_i^{\phi} \frac{\phi}{v} \right),
\label{eq:sub.H}
\end{equation}
where $v=246$~Ge$\!$V is the Higgs vacuum expectation value in the $\cal SM$
and $g_i^{\phi}$ are real numbers,
which are listed for the neutral scalar Higgs bosons of the
$\cal SM$ and the $\cal MSSM$ in Table \ref{tb:coup}.
As usual, $\alpha$ is the mixing angle between the
original neutral scalar Higgs fields of definite weak hypercharge
and the mass eigenstates, $h$ and $H$,
and $\tan\beta$ is the ratio of the vacuum expectation values of
the two Higgs doublets in the $\cal MSSM$.
In the ${\cal SM}$, the sum in Eq.~(\ref{eq:sub.H}) collapses to one item,
with $g_i^{\phi}$ being equal to unity.

\begin{table}[hbt]
\renewcommand{\arraystretch}{1.5}
\begin{center}
\begin{tabular}{|lc||ccc|} \hline
\multicolumn{2}{|c||}{$\phi$} & $t$ & $b$ & $V=W,Z$ \\ \hline \hline
${\cal SM}$ & $H$ & 1 & 1 & 1 \\ \hline
${\cal MSSM}$ & $h$ & $\cos\alpha/\sin\beta$ & $-\sin\alpha/\cos\beta$ &
$\sin(\beta-\alpha)$ \\
& $H$ & $\sin\alpha/\sin\beta$ & $\cos\alpha/\cos\beta$ &
$\cos(\beta-\alpha)$ \\ \hline
\end{tabular} \\[0.3cm]
\renewcommand{\arraystretch}{1.2}
\caption[ ]{\label{tb:coup}
\it Values of $g_i^{\phi}$ in Eq.~(\ref{eq:sub.H}) for the neutral scalar Higgs
bosons of the $\cal SM$ and the $\cal MSSM$.}
\end{center}
\end{table}

In higher orders of the perturbative expansion, the masses $m_i$, the Higgs
fields $\phi$, the couplings $g_i^{\phi}$, and the vacuum expectation value
$v$ have to be replaced by their bare counterparts, which we shall
label with the superscript 0.
This leads to the following LET
for neutral scalar Higgs bosons \citer{hggful,lhzz2}:
\begin{equation}
\lim_{p_\phi\to 0} {\cal M}(X\phi) = \sum_{i=f,V}\frac{g_{i}^{\phi0}}{v^0}\,
\frac{m_i^0\partial}{\partial m_i^0} {\cal M} (X),
\label{eq:let.H}
\end{equation}
where the symbol ${\cal M}(X)$ denotes the matrix element of any particle
configuration $X$, expressed in terms of bare quantities,
and ${\cal M}(X\phi)$ is the corresponding one with a neutral scalar
Higgs boson $\phi$ attached as an external particle in all possible ways.
The renormalization of the bare quantities is
performed after evaluating the right--hand side of Eq.~(\ref{eq:let.H}).
It is important to notice that the differentiation in Eq.~(\ref{eq:let.H})
only acts on the bare masses appearing in the propagators of the massive
particles, while bare mass--dependent couplings must be treated as constants.
The reason is that such couplings may be considered as being generated
by a substitution similar to Eq.~(\ref{eq:sub.H}), so that further
application of Eq.~(\ref{eq:sub.H}) would introduce tree--level vertices
between the Higgs bosons and the massive particles which are absent in the
$\cal SM$ and the $\cal MSSM$.

\subsection{Pseudoscalar Higgs Bosons}
%           =========================
The pseudoscalar Higgs boson $A$ of the ${\cal MSSM}$ does not interact
with the gauge bosons at tree level.
The Lagrangian for its interaction with the massive fermions reads
\begin{eqnarray}
{\cal L}_{Af\bar f} & = & -\sum_{f} m_f^0 \bar f_L^0 f_R^0 \left( 1+i g_f^{A0}
\frac{A^0}{v^0} \right) +\mbox{h.c.} \nonumber \\ \nonumber \\
& = & -\sum_f m_f^0 \left( \bar f^0 f^0 + ig_f^{A0} \bar f^0 \gamma_5 f^0
\frac{A^0}{v^0} \right),
\label{eq:Lint.A}
\end{eqnarray}
where $g_b^{A0}=1/g_t^{A0}=\tan\beta$.
{}From Eq.~(\ref{eq:Lint.A}) it is obvious that the tree--level interaction of
$A$ with the massive fermions may be generated from their mass terms by the
substitution
\cite{asubst}
\begin{displaymath}
m_f^0 \to m_f^0 \left( 1+i g_f^{A0} \frac{A^0}{v^0} \right)
\end{displaymath}
prior to adding the complex conjugate of the chiral mass operator $m_f^0 \bar
f_L^0 f_R^0$. This can be achieved more systematically by introducing left--
and right--handed masses, $m_{f\pm}^0$, and writing the bare fermion
propagators as
\begin{equation}
S_F(p) = \frac{\not\!p + m_{f+}^0 \omega_+ + m_{f-}^0 \omega_-}
{p^2 - m_{f+}^0 m_{f-}^0},
\end{equation}
where $\omega_\pm = (1\pm \gamma_5)/2$ are the chiral projectors.
Then, the $Af\bar f$ interaction
may be generated by the substitutions
\begin{equation}
\left. m_{f\pm}^0 \to m_{f\pm}^0 \left( 1\pm ig_f^{A0} \frac{A^0}{v^0} \right)
\right|_{m_{f\pm}^0 = m_f^0}.
\label{eq:sub.A}
\end{equation}
This leads to the following LET for the $Af\bar f$ interaction:
\begin{equation}
\left.
\lim_{p_A\to 0} {\cal M}(XA) = \sum_f i\frac{g_f^{A0}}{v^0} m_f^0\left(
\frac{\partial}{\partial m_{f+}^0} - \frac{\partial}{\partial m_{f-}^0} \right)
{\cal M}(X)
\right|_{m_{f\pm}^0 = m_f^0},
\label{eq:let.a}
\end{equation}
where ${\cal M}(X)$ denotes the matrix element of any particle configuration
$X$ and ${\cal M}(XA)$ is the corresponding one with an external pseudoscalar
Higgs boson $A$ added in all possible ways. Again, the renormalization of the
bare quantities is to be performed after the right--hand side of
Eq.~(\ref{eq:let.a}) has been evaluated.

However, substitution (\ref{eq:sub.A}) does not yield the full
effective Lagrangian.
In the case of the interaction of a pseudoscalar particle with vector bosons,
additional contributions may arise due to the Adler--Bell--Jackiw (ABJ) anomaly
\cite{abj}.
Such contributions appear if an odd number of external pseudoscalar Higgs
bosons, which carry odd ${\cal CP}$ parity at vanishing momentum transfer,
is coupled to a pair of vector bosons via a single fermion loop.
Therefore, the LETs for odd numbers of external pseudoscalar Higgs particles
differ from those for even numbers.

\subsubsection{Odd Number of Pseudoscalars}
%              ===========================
In the case of an odd number of pseudoscalar Higgs bosons coupled to one
heavy--fermion loop, a contribution related the ABJ anomaly has to be added
to the effective Lagrangian of the model with the heavy fermion integrated out.
This contribution may be derived by observing that,
in addition to the pseudoscalar mass term,
the divergence of the axial vector current, $j_5^\mu = \bar f \gamma^\mu
\gamma_5 f$, receives a contribution from the ABJ anomaly \cite{abj},
\begin{equation}
\partial_\mu j_5^\mu = 2i m_f \bar f \gamma_5 f + \sum_{V,V'=g,\gamma,W,Z}
\frac{\alpha_{VV'}} {4\pi} V^{a\mu\nu} \widetilde{V}'^a_{\mu\nu},
\label{eq:ABJ}
\end{equation}
where $V^{a\mu\nu}$ is the field--strength tensor of $V$ and
$\widetilde{V}^a_{\mu\nu} = \epsilon_{\mu\nu\rho\sigma} V^{a\rho\sigma}$
is its dual.
Here, the index $a$ stems from the respective gauge group.
The couplings $\alpha_{VV'}$ are listed for the $\cal SM$ (and the
$\cal MSSM$) gauge bosons in Table \ref{tb:alphaVV'}.
As usual, $\alpha_{em}$ is the fine--structure constant,
$G_F$ is the Fermi constant,
$v_f = 2I_{3f}-4e_fs_w^2$, $a_f=2I_{3f}$, $c_w^2=1-s_w^2=M_W^2/M_Z^2$, and
$N_{c}$, $e_f$, and $I_{3f}$ are the number of colours, the fractional charge,
and the third isospin component of the fermion $f$, respectively.

\begin{table}[hbt]
\renewcommand{\arraystretch}{2.0}
\begin{center}
\begin{tabular}{|c||c|c|c|c|} \hline
$\alpha_{VV'}$ & $g$ & $\gamma$ & $Z$ & $W$ \\ \hline \hline
$g$ & $\alpha_s$ & 0 & 0 & 0 \\ \hline
$\gamma$ & 0 & $N_{c}e_f^2 \alpha_{em}$ & $N_{c}e_f v_f
\sqrt{\frac{\alpha_{em}G_FM_Z^2}{8\sqrt{2}\pi}}$ & 0 \\ \hline
$Z$ & 0 & $N_{c}e_f v_f \sqrt{\frac{\alpha_{em}G_FM_Z^2}{8\sqrt{2}\pi}}$ &
$N_{c}\frac{G_FM_Z^2}{8\sqrt{2}\pi}~\left( v_f^2+ \frac{a_f^2}{3} \right)$
& 0 \\ \hline
$W$ & 0 & 0 & 0 & $N_{c}\frac{G_F M_W^2}{2\sqrt{2}\pi}$ \\ \hline
\end{tabular}
\renewcommand{\arraystretch}{1.2}
\caption[ ]{\label{tb:alphaVV'}\it Values of $\alpha_{VV'}$ in
Eq.~(\ref{eq:ABJ}) for the $\cal SM$ (and the $\cal MSSM$) gauge bosons
$V,V'$.}
\end{center}
\end{table}

The Adler--Bardeen theorem \cite{adlbar} states that Eq.~(\ref{eq:ABJ}) is not
modified by radiative corrections.
On the other hand, we have \cite{sutvel}
\begin{equation}
\lim_{p_A\to 0} \langle A| \partial_\mu j_5^\mu A |VV' \rangle = 0.
\end{equation}
This allows us to derive from Eq.~(\ref{eq:Lint.A}) the anomalous part of the
low--$p_A$ effective interaction Lagrangian of $A$ \citer{labj,labj3},
\begin{equation}
{\cal L}_{ABJ} = g_f^A \sum_{V,V'} \frac{\alpha_{VV'}}{8\pi} V^{a\mu\nu}
\widetilde{V}'^a_{\mu\nu} \frac{A}{v},
\label{eq:L.AVV}
\end{equation}
which is valid to all orders.
This contribution has to be added to the part of the effective Lagrangian
which is generated by LET~(\ref{eq:let.a}).

\subsubsection{Even Number of Pseudoscalars}
%              ============================
An even number of pseudoscalar Higgs bosons carry positive ${\cal CP}$ parity
at
vanishing momentum transfer. Hence, the ABJ anomaly does not contribute to
the effective Lagrangian for an even number of pseudoscalars coupled to a
single heavy--fermion loop.
Substitution (\ref{eq:sub.A}) leads us to the following LET:
\begin{equation}
\left.
\lim_{p_A\to 0} {\cal M}(XA^{2n}) = \sum_{f} \left( i\frac{g_f^{A0}}{v^0}
\right)^{2n} (m_f^0)^{2n} \left( \frac{\partial}{\partial m_{f+}^0} -
\frac{\partial}{\partial m_{f-}^0} \right)^{2n} {\cal M}(X)
\right|_{m_{f\pm}^0 = m_f^0}.
\label{eq:let.A}
\end{equation}
As in the scalar case,
the renormalization of the bare quantities
has to be performed after taking the derivative on the right--hand side
of Eq.~(\ref{eq:let.A}), and
mass--dependent couplings must be kept fixed with respect to mass
differentiation.
Notice that we may also use Eq.~(\ref{eq:let.A}) for odd numbers of $A$
bosons, if we take $X$ to implicitly include one of them.

\section{Applications at One Loop}
%        ========================
In the following, we shall consider generic neutral scalar and pseudoscalar
Higgs bosons, $H$ and $A$, with $g_t^H=g_V^H=g_t^A=1$ and $g_V^A=0$ ($V=W,Z$).
For simplicity, we shall neglect the masses of all loop fermions, except for
the top quark.
Our results can easily be generalized to arbitrary couplings by means of
Eqs.~(\ref{eq:sub.H}) and (\ref{eq:sub.A}).

\subsection{Higgs Couplings to Two Photons and Two Gluons}
%           =============================================
\subsubsection{Scalar Higgs Bosons}
%              ===================
In order to calculate the effective coupling of the neutral scalar Higgs boson
$H$ to two photons, we have to evaluate the contributions from the charged
massive particles, {\it i.e.}, in our case the top quark and the $W$ boson,
to the on--shell photon self--energy.
The result may be formulated as the following effective Lagrangian:
\begin{equation}
{\cal L}_{\gamma\gamma} = -\frac{1}{4} F^{0\mu\nu} F_{\mu\nu}^0 \left[ 1+
\Pi_{\gamma\gamma}^t(0) + \Pi_{\gamma\gamma}^W(0)\right],
\label{eq:Lga.eff}
\end{equation}
where $F_{\mu\nu}^0$ is the bare electromagnetic field--strength tensor and
\begin{eqnarray}
\label{gagat}
\Pi_{\gamma\gamma}^t(0) & = &
 N_c e_t^2 \frac{\alpha_{em}}{3\pi}
 \left(\frac{4\pi\mu^2}{m_t^2}\right)^\epsilon\Gamma(1+\epsilon)
\frac{1}{\epsilon}, \\
\nonumber \\
\label{gagaw}
\Pi_{\gamma\gamma}^W(0) & = &
-\frac{\alpha_{em}}{4\pi}
\left(\frac{4\pi\mu^2}{M_W^2}\right)^\epsilon\Gamma(1+\epsilon)
\left[\frac{7}{\epsilon}+\frac{2}{3}+{\cal O}(\epsilon)\right]
\end{eqnarray}
are the lowest--order expressions of the top--quark and $W$--boson
contributions to the (dimensionless) photon vacuum--polarization function at
zero momentum transfer, respectively.
Notice that Eq.~(\ref{gagaw}) has been calculated using the pinch technique
\cite{pt} and is thus manifestly gauge independent.
Application of LET (\ref{eq:let.H}) leads to \cite{hsubst1,hsubst2}
\begin{equation}
\label{eq:L.HgagaLO}
{\cal L}_{H\gamma\gamma} = \frac{\alpha_{em}}{2\pi}
F^{\mu\nu} F_{\mu\nu} \frac{H}{v}
\left( N_c\frac{e_t^2}{3} - \frac{7}{4} \right),
\end{equation}
which is valid for $M_H\ll M_W,m_t$.
This expression is in agreement with the leading term of
the full one--loop result \cite{hsubst1,hsubst2}.\footnote{If
we use an ultraviolet cut--off, $\Lambda_{UV}$, instead of dimensional
regularization, then Eq.~(\ref{gagat}) assumes the form
$\Pi_{\gamma\gamma}^t(0) = N_c e_t^2 (\alpha_{em}/3\pi)
\log( \Lambda_{UV}^2 / m_t^2)$ \cite{theorem2}, which also leads to the first
term of Eq.~(\ref{eq:L.HgagaLO}). This nicely demonstrates that this term
is independent of the regularization scheme.}
We may upgrade the range of validity of Eq.~(\ref{eq:L.HgagaLO}) to be
$M_H\ll m_t$ by replacing the second term contained within the parentheses
with the full $M_W$--dependent result \cite{hsubst2}.

The effective Lagrangian ${\cal L}_{H\gamma\gamma}$ of
Eq.~(\ref{eq:L.HgagaLO}) fixes the photonic Higgs decay width
$\Gamma (H\to\gamma\gamma )$ as well as the cross section of Higgs
production via photon fusion $\sigma(\gamma\gamma\to H)$.
For $M_H\lsim140$~Ge$\!$V, the decay $H\to \gamma\gamma$ has a branching ratio
of order $10^{-3}$ and will play an important r\^ole for the search for the
Higgs boson in this mass range at the LHC \cite{hgamlhc}.
On the other hand, $\gamma\gamma\to H$ will be the relevant
Higgs--boson--production mechanism at future photon colliders \cite{gamfus}.

Similarly to the $H\gamma\gamma$ case,
the derivation of the effective $Hgg$ Lagrangian starts from
\begin{equation}
{\cal L}_{gg}=-\frac{1}{4}G^{a\mu\nu}G_{\mu\nu}^a\left[1+\Pi_{gg}^t(0)\right],
\end{equation}
where
\begin{equation}
\label{piggt}
\Pi_{gg}^t(0) = \frac{\alpha_s}{6\pi}
\left[\frac{4\pi\mu^2}{(m_t^0)^2}\right]^\epsilon\Gamma(1+\epsilon)
\frac{1}{\epsilon}
\end{equation}
is the top--quark contribution to the (dimensionless) gluon self--energy,
and yields \cite{hsubst1}
\begin{equation}
{\cal L}_{Hgg} = \frac{\alpha_s}{12\pi} G^{a\mu\nu} G^a_{\mu\nu}
\frac{H}{v},
\label{eq:L.HggLO}
\end{equation}
which is valid for $M_H\ll m_t$.
This Lagrangian determines the gluonic decay width $\Gamma (H\to gg)$,
which, for $M_H\lsim150$~Ge$\!$V, has a branching ratio of a few percent
and should be observable at future $e^+e^-$ colliders \cite{eehgg}.
Furthermore, it controls the production of a light Higgs boson via gluon
fusion $gg\to H$, which will be the dominant production mechanism of this
particle at the LHC \cite{hgamlhc,glufus}.

\subsubsection{Pseudoscalar Higgs Bosons}
%              =========================
Due to the absence of a tree--level $AWW$ vertex and the fact that
Eq.~(\ref{eq:let.a}) annihilates the top--quark contributions
to the photon and gluon self--energies,
the effective $A\gamma\gamma$ and $Agg$ Lagrangians only consist of the
ABJ parts of Eq.~(\ref{eq:L.AVV}) \cite{labj,labj2},
\begin{eqnarray}
{\cal L}_{A\gamma\gamma} & = &  N_c e_t^2\frac{\alpha_{em}}{8\pi} F^{\mu\nu}
\widetilde{F}_{\mu\nu} \frac{A}{v}, \label{eq:L.Agaga} \\ \nonumber \\
{\cal L}_{Agg} & = & \frac{\alpha_s}{8\pi} G^{a\mu\nu}
\widetilde{G}^a_{\mu\nu} \frac{A}{v}.
\label{eq:L.Agg}
\end{eqnarray}
We recall that these effective Lagrangians do not receive radiative
corrections.
They determine the photonic and gluonic decays of a pseudoscalar Higgs boson
$A$, with $M_A\ll m_t$, as well as its single production via photon and gluon
fusion, which will be the dominant production mechanisms at future photon
colliders and the LHC \cite{hgamlhc,glufus}, respectively.

\subsection{Multi--Higgs Couplings to Two Photons and Two Gluons}
%           ====================================================
\subsubsection{Scalar Higgs Bosons}
%              ===================
We may derive an effective Lagrangian describing the coupling of any number of
neutral scalar Higgs bosons $H$ to two photons by iteratively applying
LET~(\ref{eq:let.H}) to Eq.~(\ref{eq:L.HgagaLO}), which we rewrite as
\begin{equation}
{\cal L}_{H^1\gamma\gamma} =\frac{\alpha_{em}}{2\pi} F^{\mu\nu} F_{\mu\nu} H
\left(N_c\frac{e_t^2}{3}\,\frac{g_t}{m_t} - \frac{7}{4}\,\frac{g_W}{M_W}
\right),
\end{equation}
since the coupling constants of the Higgs boson to the top--quark and the
$W$--boson,  $g_t = m_t/v$ and $g_W = M_W/v$, have to be kept fixed with
respect to the mass differentiation.
The effective Lagrangian for the interactions of $n$ $H$ bosons with two
photons is then given by
\begin{eqnarray}
{\cal L}_{H^n\gamma\gamma} & = &  F^{\mu\nu} F_{\mu\nu}\frac{H^n}{n!}\,
\frac{\alpha_{em}}{2\pi}
\left(N_c\frac{e_t^2}{3}g_t^n \frac{\partial^{n-1}}{\partial m_t^{n-1}}\,
\frac{1}{m_t} - \frac{7}{4} g_W^n \frac{\partial^{n-1}}{\partial M_W^{n-1}}\,
\frac{1}{M_W} \right) \nonumber \\ \nonumber \\
& = & - F^{\mu\nu} F_{\mu\nu} \frac{1}{n} \left( -
\frac{H}{v} \right)^n\frac{\alpha_{em}}{2\pi}
\left(N_c\frac{e_t^2}{3} - \frac{7}{4} \right).
\end{eqnarray}
Summing up all these Lagrangians, we find \cite{hsubst2}
\begin{equation}
{\cal L}_{H\gamma\gamma} = \sum_{n=1}^\infty {\cal L}_{H^n\gamma\gamma} =
\frac{\alpha_{em}}{2\pi}
\left( N_c \frac{e_t^2}{3} - \frac{7}{4} \right)
F^{\mu\nu} F_{\mu\nu} \log\left( 1+ \frac{H}{v} \right).
\end{equation}
The analogous calculation for the gluon case yields \cite{hagiwa}
\begin{equation}
{\cal L}_{Hgg} = \frac{\alpha_s}{12\pi} G^{a\mu\nu} G^a_{\mu\nu}
\log\left( 1+ \frac{H}{v} \right).
\end{equation}
These Lagrangians govern the cross sections for multi--Higgs production via
photon and gluon fusion in the limit $M_H\ll m_t$, where Higgs
self--interactions are suppressed.

\subsubsection{Pseudoscalar Higgs Bosons}
%              =========================
Similarly to the scalar case, the effective Lagrangian describing the
coupling of any number of pseudoscalar Higgs bosons $A$ to two photons (gluons)
can be deduced by iterative application of LET (\ref{eq:let.A}).
For even numbers of pseudoscalar Higgs bosons, LET (\ref{eq:let.A}) has to be
applied to the photon (gluon) self--energy, while for odd numbers it has to be
applied to Lagrangian (\ref{eq:L.Agaga}) [(\ref{eq:L.Agg})].
In this way, we obtain
\begin{eqnarray}
{\cal L}_{A^{2n}\gamma\gamma} & = & \left.
- N_c e_t^2\frac{\alpha_{em}}{12\pi} F^{\mu\nu}
F_{\mu\nu} \frac{A^{2n}}{(2n)!} \left[ i\frac{m_t}{v} \left( \frac{\partial}
{\partial m_{t+}} - \frac{\partial}{\partial m_{t-}} \right)\right]^{2n} \log
(m_{t+} m_{t-})
\right|_{m_{t\pm} = m_t},
\nonumber \\ \nonumber \\
{\cal L}_{A^{2n+1}\gamma\gamma} & = & \left.
-i N_c e_t^2 \frac{\alpha_{em}}{16\pi} F^{\mu\nu}\widetilde{F}_{\mu\nu}
\frac{A^{2n+1}}{(2n+1)!} \left[ i \frac{m_t}{v}\left(
\frac{\partial}{\partial m_{t+}} - \frac{\partial}{\partial m_{t-}} \right)
\right]^{2n+1} \log \frac{m_{t+}}{m_{t-}}
\right|_{m_{t\pm} = m_t},
\nonumber \\ \nonumber \\
&&
\end{eqnarray}
and similarly for gluons.
Summing separately over even and odd numbers of pseudoscalar Higgs bosons,
we obtain
\begin{eqnarray}
{\cal L}_{A\gamma\gamma}^{(even)} & = & N_ce_t^2\frac{\alpha_{em}}{12\pi}
F^{\mu\nu}F_{\mu\nu}
\log\left( 1+ \frac{A^2}{v^2} \right), \\ \nonumber \\
{\cal L}_{Agg}^{(even)} & = & \frac{\alpha_s}{24\pi} G^{a\mu\nu} G^a_{\mu\nu}
\log\left( 1+ \frac{A^2}{v^2} \right)
\label{eq:L.Angg}
\end{eqnarray}
for even numbers and
\begin{eqnarray}
{\cal L}_{A\gamma\gamma}^{(odd)} & = & - i N_c e_t^2 \frac{\alpha_{em}}{16\pi}
F^{\mu\nu}
\widetilde{F}_{\mu\nu} \left[\log\left( 1+ i \frac{A}{v}\right) -
\log\left( 1- i \frac{A}{v} \right) \right],\nonumber \\ \nonumber \\
{\cal L}_{Agg}^{(odd)} & = & - i \frac{\alpha_s}{16\pi} G^{a\mu\nu}
\widetilde{G}^a_{\mu\nu} \left[\log\left( 1+ i \frac{A}{v}\right) -
\log\left( 1- i \frac{A}{v} \right) \right]
\end{eqnarray}
for odd numbers.
Lagrangian (\ref{eq:L.Angg}) agrees with the pseudoscalar $\chi^3$ part
of Ref.~\cite{hagiwa}.
These Lagrangians describe the production of many pseudoscalar Higgs bosons
by photon and gluon fusion, which may be relevant at future photon
colliders and the LHC, respectively.

\subsection{Higgs Couplings to One $Z$ Boson and One Photon}
%           ===============================================
\subsubsection{Scalar Higgs Bosons}
%              ===================
In Section~3.1.1, we have seen how the
$H\gamma\gamma$ coupling is related to the photon self--energy.
In a similar fashion, the
$HZ\gamma$ coupling may be derived from the $\gamma$--$Z$
transition amplitude.
In order for the $HZ\gamma$ amplitude to be gauge independent, all three
external particles must be on their mass shells \cite{hzga}.
Then, however, it is unjustified to integrate out the virtual $W$ boson.
In the following, we shall therefore concentrate on the top--quark loop.
Similarly to Eq.~(\ref{eq:Lga.eff}), we may write
\begin{equation}
{\cal L}_{Z\gamma} = -\frac{1}{4} F^{0\mu\nu} Z_{\mu\nu}^0
\Pi_{Z\gamma}^t(0),
\label{eq:LZga.eff}
\end{equation}
where $Z_{\mu\nu}^0$ is the bare $Z$--boson field--strength tensor and
\begin{equation}
\Pi_{Z\gamma}^t(0)  =  \frac{N_ce_tv_t}{3\pi}
\sqrt{\frac{\alpha_{em}G_FM_Z^2}{8\sqrt2\pi}}
\left(\frac{4\pi\mu^2}{m_t^2}\right)^\epsilon\Gamma(1+\epsilon)
\frac{1}{\epsilon}
\end{equation}
is the top--quark contribution to the (dimensionless)
$Z$--$\gamma$ mixing amplitude at zero momentum transfer.
Here, we have used the notation introduced below Eq.~(\ref{eq:ABJ}).
Differentiating this expressions with respect to $m_t$,
we end up with the effective $HZ\gamma$ Lagrangian,
\begin{equation}
{\cal L}_{HZ\gamma} =\frac{N_ce_tv_t}{6\pi}
\sqrt{\frac{\alpha_{em}G_FM_Z^2}{8\sqrt2\pi}}
F^{\mu\nu} Z_{\mu\nu} \frac{H}{v},
\label{eq:L.HZgaLO}
\end{equation}
appropriate to the limit where $M_H,M_Z\ll m_t$.
Lagrangian (\ref{eq:L.HZgaLO}) may also be derived by directly expanding
the corresponding one--loop diagram \cite{hzga,hzga2}.
The $W$--boson contribution may be found in Ref.~\cite{hzga}.
It is significant and must be included in order to obtain a satisfactory
description \cite{hzga,hzga2}.
The full $HZ\gamma$ Lagrangian determines the width of the rare $Z$--boson
decay $Z\to H\gamma$.

\subsubsection{Pseudoscalar Higgs Bosons}
%              =========================
Similarly to the $A\gamma\gamma$ case discussed in Section~3.1.2,
the coupling of the pseudoscalar Higgs boson $A$ to a $Z$ boson and a photon is
controlled by Lagrangian (\ref{eq:L.AVV}). Specifically, we have
\begin{equation}
{\cal L}_{AZ\gamma} = \frac{N_c e_t v_t}{4\pi}\sqrt{\frac{\alpha_{em}G_F M_Z^2}
{8\sqrt{2}\pi}}Z^{\mu\nu} \widetilde{F}_{\mu\nu}\frac{A}{v}.
\label{eq:L.AZga}
\end{equation}
This Lagrangian fixes the width of the rare $Z\to A\gamma$ decay
for $M_A,M_Z\ll m_t$.

\subsection{$H\to b \bar b$}
%           ===============
By means of LET (\ref{eq:let.H}), we may also extract the ${\cal O}(G_F m_t^2)$
correction to the $b\bar b$ decay rate of the neutral scalar Higgs bosons $H$
in the limit where $M_H\ll m_t$ \cite{hbb2}.
This just requires knowledge of the ${\cal O}(G_Fm_t^2)$ contribution to the
bottom--quark self--energy.
To compute this contribution, we must put the bottom quark on mass shell and
neglect its mass, except for one overall power.
Furthermore, it is sufficient to take into account the longitudinal component
of the $W$ boson, $w^\pm$, which we may take to be massless, too.
The bare amplitude describing the propagation of the $b$ quark
can be cast into the form
\begin{equation}
{\cal M} (b\to b) =\bar b^0\left\{
 m_b^0 \left[-1+\Sigma_S(0) \right] + \not \! p \left[
\Sigma_V (0) + \gamma_5 \Sigma_A (0) \right]\right\}b^0.
\label{eq:btob}
\end{equation}
The Yukawa couplings of $w^\pm$ to the bottom and top quarks must be kept
fixed with respect to the mass differentiation.
Therefore, we call them $g_q^0 = m_q^0 / v^0$ ($q=t,b$).
The various amplitudes in Eq.~(\ref{eq:btob}) are, at ${\cal O}(G_Fm_t^2)$,
given by \cite{hbb1}
\begin{eqnarray}
\label{bsigs}
m_b^0 \Sigma_S (0) & = & -\frac{g_b^0g_t^0}{(4\pi)^2} m_t^0
\left[ \frac{4\pi \mu^2}{(m_t^0)^2}\right]^\epsilon
\Gamma(1+\epsilon)\left[ \frac{2}{\epsilon}
+ 2 + 2\epsilon +{\cal O}(\epsilon^2)\right], \\ \nonumber \\
\label{bsigv}
\Sigma_V (0) & = & \frac{(g_t^0)^2}{(4\pi)^2}
\left[ \frac{4\pi \mu^2}{(m_t^0)^2}\right]^\epsilon\Gamma(1+\epsilon) \left[
\frac{1}{2\epsilon} + \frac{3}{4} + \frac{7}{8}\epsilon+{\cal O}(\epsilon^2)
\right], \\ \nonumber \\
\Sigma_A (0) & = & -\Sigma_V (0).
\end{eqnarray}
Applying LET (\ref{eq:let.H}) to Eq.~(\ref{eq:btob}),
\begin{equation}
\lim_{p_H\to 0} {\cal M} (b\to bH) = \frac{1}{v^0} \left( \frac{m_b^0\partial}
{\partial m_b^0} + \frac{m_t^0\partial}{\partial m_t^0} \right) {\cal M}
(b\to b),
\end{equation}
and then using the Dirac equation, we find the following effective
Lagrangian:
\begin{equation}
\label{lhbb}
{\cal L}_{Hb\bar b} = -m_b^0 \bar b^0 b^0 \frac{H^0}{v^0}(1
+\delta_{Hb\bar b}^0),
\end{equation}
where
\begin{eqnarray}
\label{delhbbzero}
\delta_{Hb\bar b}^0 & = & -m_t^0 \frac{\partial[\Sigma_S(0) +
\Sigma_V (0)]}{\partial m_t^0}\nonumber\\ \nonumber\\
& = & -\Sigma_S (0) +2\epsilon [\Sigma_S (0) + \Sigma_V (0) ].
\end{eqnarray}
It should be noted that the axial part $\Sigma_A (0)$ is eliminated by
the Dirac equation.  Next, we have to renormalize the bottom--quark
mass and wave function.
This may be achieved by substituting
\begin{equation}
\label{brenor}
m_b^0\bar b^0b^0=m_b\bar bb\frac{1}{1-\Sigma_S(0)}.
\end{equation}
In this way, we obtain the non--universal correction to the effective
$Hb\bar b$ Lagrangian \cite{hbb2,hbb1},
\begin{equation}
{\cal L}_{Hb\bar b} = -m_b \bar b b \frac{H^0}{v^0} ( 1+
\delta_{Hb\bar b,\,nu}),
\end{equation}
with
\begin{equation}
\label{deltahbb}
\delta_{Hb\bar b,\,nu} = 2\epsilon [\Sigma_S (0) + \Sigma_V (0) ] = -3x_t^0,
\end{equation}
where $x_t^0 = G_F (m_t^0)^2 /(8\sqrt{2}\pi^2)$.
Using also the universal relation \cite{hbb2,hbb1}
\begin{equation}
\frac{H^0}{v^0}=2^{1/4}G_F^{1/2}H(1+\delta_u),
\end{equation}
with
\begin{equation}
\label{deltau}
\delta_u  =  \frac{7}{6}N_c x_t,
\end{equation}
we find
\begin{equation}
\label{hbbeff}
{\cal L}_{Hb\bar b} = -2^{1/4}G_F^{1/2}m_b \bar b b H
( 1+\delta_{Hb\bar b} ),
\end{equation}
with
\begin{equation}
\label{delhbb}
\delta_{Hb\bar b}=\delta_{Hb\bar b,\,nu}+\delta_u
=\left(\frac{7}{6}N_c-3\right)x_t,
\end{equation}
where we have used that $m_t^0$ and $m_t$ coincide in lowest order.
The ${\cal O} (G_F m_t^2)$ correction to $\Gamma (H\to b \bar b)$ is then
$2\delta_{Hb\bar b}$, which agrees with the explicit calculation \cite{hbb1}.
For $m_t=176$~Ge$\!$V, this term enhances $\Gamma(H\to b\bar b)$ by
approximately 0.3\%, but it does not yet dominate the full weak correction at
one loop.
For example, at $M_H=70$~Ge$\!$V, the latter amounts to approximately
$-0.4\%$.

\subsection{$HZZ$ coupling and $e^+e^- \to ZH$}
%           =============================
Another useful application of LET~(\ref{eq:let.H})
is to derive the leading top--mass dependent correction of
${\cal O}(G_Fm_t^2)$ to the coupling of the neutral scalar boson $H$ to a
pair of $Z$ bosons.
The starting point is the amplitude describing the propagation of an
on--shell $Z$ boson interacting with virtual top quarks,
\begin{equation}
{\cal M}(Z\to Z)=\frac{1}{2}Z^{0\mu}Z_\mu^0\left[(M_Z^0)^2-\Pi_{ZZ}(0)\right],
\label{eq:MZZ}
\end{equation}
where \cite{pizz1,hzzful1}
\begin{equation}
\Pi_{ZZ} (0)  =  -2N_c(g_Z^0v^0)^2x_t^0
\left[\frac{4\pi\mu^2}{(m_t^0)^2}\right]^\epsilon
\Gamma(1+\epsilon)\frac{1}{\epsilon}
\end{equation}
is the top--quark contribution to the $Z$--boson self--energy.
Here, $g_Z^0 = M_Z^0/v^0$, $x_t^0$ is defined below Eq.~(\ref{deltahbb}), and
we have used $M_Z\ll m_t$ in the loop amplitude.
In the case at hand, LET~(\ref{eq:let.H}) takes the form
\begin{equation}
\lim_{p_H\to 0} {\cal M} (Z\to ZH) = \frac{1}{v^0} \left(
\frac{m_t^0\partial}{\partial m_t^0}+
\frac{M_Z^0\partial}{\partial M_Z^0}\right){\cal M}(Z\to Z),
\label{eq:MHZZ}
\end{equation}
where $g_Z^0$ has to be kept fixed.
After evaluating the right--hand side of Eq.~(\ref{eq:MHZZ}), we are in a
position to write down the effective $HZZ$ Lagrangian,
\begin{equation}
{\cal L}_{HZZ} = (M_Z^0)^2 Z^{0\mu} Z_\mu^0 \frac{H^0}{v^0}
(1+\delta_{HZZ}^0),
\end{equation}
where
\begin{equation}
\delta_{HZZ}^0 = -(1-\epsilon) \frac{\Pi_{ZZ} (0)}{(M_Z^0)^2}.
\end{equation}
Renormalizing the $Z$--boson mass and wave function,
\begin{eqnarray}
(M_Z^0)^2 & = & M_Z^2 + \delta M_Z^2, \nonumber \\ \nonumber \\
Z_\mu^0  & = &  Z_\mu ( 1+\delta Z_Z )^{1/2},
\end{eqnarray}
with the on--shell counter terms
\begin{eqnarray}
\delta M_Z^2 & = & \Pi_{ZZ} (0), \nonumber \\ \nonumber \\
\delta Z_Z & = & - \Pi_{ZZ}^\prime(0),
\end{eqnarray}
we obtain the finite non--universal correction to the $HZZ$
coupling,
\begin{equation}
\delta_{HZZ,\,nu} = \epsilon\frac{\Pi_{ZZ} (0)}{(M_Z^0)^2}=-2N_cx_t^0.
\end{equation}
It should be noted that $\delta Z_Z$ does not receive any
contribution in ${\cal O}(G_F m_t^2)$ and thus does not contribute here.
Replacing the bare Higgs field $H^0$ and the bare vacuum expectation value
$v^0$ with their renormalized counterparts,
we introduce the universal correction $\delta_u$ of Eq.~(\ref{deltau}).
Consequently, the effective Lagrangian reads \cite{hzzful1}
\begin{equation}
{\cal L}_{HZZ}  =  2^{1/4}G_F^{1/2} M_Z^2 Z^\mu Z_\mu H
(1+\delta_{HZZ}),
\label{eq:LHZZ1}
\end{equation}
with
\begin{equation}
\label{deltahzz}
\delta_{HZZ}=\delta_{HZZ,\,nu}+\delta_u=-\frac{5}{6}N_cx_t.
\end{equation}
The decay width $\Gamma(H\to ZZ)$ is then corrected by the factor
$(1+2\delta_{HZZ})$,
which agrees with the expansion of the full one--loop correction
\cite{hzzful1}.
This provides an approximation for $2M_Z <M_H \ll m_t$,
which is not satisfied for the actual $Z$--boson and top--quark masses.
However, this result may be used as a building block for the calculation
of the ${\cal O}(G_Fm_t^2)$ correction to the Higgs production mechanism
$e^+ e^- \to ZH$.
In fact, by invoking the improved Born approximation \cite{iba} for the
on--shell scheme formulated with $G_F$, we find that $\sigma (e^+ e^- \to ZH)$
is corrected by the factor $(1+\delta_{HZe^+ e^-})$, where \cite{eehzful}
\begin{eqnarray}
\label{deltahzee}
\delta_{HZe^+ e^-} & = & 2\delta_{HZZ} + \left( 1-8 \frac{c_w^2 Q_e v_e}{v_e^2
+ a_e^2} \right) \Delta \rho\\ \nonumber\\
& = & -2N_cx_t \left(\frac{1}{3}+ 4\frac{c_w^2 Q_e v_e}{v_e^2+ a_e^2} \right),
\end{eqnarray}
with $\Delta \rho = N_cx_t$ \cite{drho1}.
This agrees with the corresponding expansion of the full one--loop correction
\cite{eehzful}.
Numerically, we find $\delta_{HZe^+ e^-}\approx-1\%$ for $m_t=176$~Ge$\!$V.
This has to be compared with the full one--loop correction, which, for
$M_H=70$~Ge$\!$V and LEP2 energy, amounts to approximately $-3\%$.

\subsection{$HWW$ coupling and $e^+e^- \to \nu_e\bar\nu_eH$}
%           ==========================================
In close analogy to the $HZZ$ case, we can also derive
the ${\cal O}(G_F m_t^2)$ correction to the $HWW$ coupling
by using LET (\ref{eq:let.H}).
Starting from the amplitude describing the propagation of an
on--shell $W$ boson in the presence of virtual top and bottom quarks,
\begin{equation}
{\cal M}(W\to W) = (W^{+\mu})^0(W_\mu^-)^0
\left[(M_W^0)^2 - \Pi_{WW}(0)\right],
\label{eq:MWW}
\end{equation}
where \cite{pizz1,hzzful1}
\begin{equation}
\Pi_{WW} (0) = -N_c(g_W^0 v^0)^2x_t^0
\left[\frac{4\pi\mu^2}{(m_t^0)^2}\right]^\epsilon
\Gamma(1+\epsilon)\left[\frac{2}{\epsilon} + 1 + \frac{\epsilon}{2}
+ {\cal O}(\epsilon^2)\right],
\end{equation}
with $g_W^0 =M_W^0/v^0$,
is the respective contribution to the $W$--boson self--energy,
applying LET~(\ref{eq:let.H}) in the form
\begin{equation}
\lim_{p_H\to 0} {\cal M} (W\to WH) = \frac{1}{v^0} \left(
\frac{m_t^0\partial}{\partial m_t^0}+\frac{M_W^0\partial}{\partial M_W^0}
\right) {\cal M}(W\to W),
\label{eq:MHWW}
\end{equation}
where $g_W^0$ must be treated as a constant,
and renormalizing the parameters according to the on--shell scheme,
we end up with the effective $HWW$ Lagrangian,
\begin{equation}
{\cal L}_{HWW}  =  2^{5/4}G_F^{1/2} M_W^2 W^{+\mu} W^-_\mu H
(1+\delta_{HWW}),
\label{eq:LHWW1}
\end{equation}
with
\begin{equation}
\label{deltahww}
\delta_{HWW}  =  -\frac{5}{6}N_cx_t,
\end{equation}
which coincides with $\delta_{HZZ}$ of Eq.~(\ref{deltahzz}).

This contains all the information which is necessary to compute the
${\cal O}(G_Fm_t^2)$ correction to Higgs--boson production via $W$--boson
fusion, $e^+e^-\to H\nu_e\bar\nu_e$, in the $G_F$ representation of the
on--shell scheme.
Since $G_F$ is defined via a charged--current process, namely
the muon decay, there are no additional ${\cal O}(G_Fm_t^2)$ corrections
from the $W$--boson propagators in this case.
Thus, the correction factor for $\sigma(e^+e^-\to H\nu_e\bar\nu_e)$ is just
$(1+2\delta_{HWW})$, which amounts to a reduction by about 2\%.

\section{Applications at Two Loops}
%        =========================
\subsection{Higgs Couplings to Two Photons}
%           ==============================
\subsubsection{Scalar Higgs Bosons}
%              ===================
In order to evaluate the two--loop QCD correction to the two--photon coupling
of the neutral scalar Higgs boson $H$ by means of the LET,
we have to start from the top--quark contribution to the photon
self--energy given in Eq.~(\ref{gagat})
and include its leading--order QCD correction,
\begin{equation}
\Pi_{\gamma\gamma}^{t(2)}(0) = N_cC_Fe_t^2
\frac{\alpha_{em}\alpha_s}{8\pi^2}
\left(\frac{4\pi\mu^2}{m_t^2}\right)^{2\epsilon}\Gamma^2(1+\epsilon)
\left[\frac{1}{\epsilon}+\frac{15}{2}+{\cal O}(\epsilon)\right],
\label{eq:PiQ.2}
\end{equation}
where $C_F=(N_c^2-1)/(2N_c)=4/3$, $m_t$ denotes the on--shell mass of the top
quark, and it is understood that Eq.~(\ref{gagat}) is also written with $m_t$.
LET (\ref{eq:let.H}) leads us to differentiate $\Pi_{\gamma\gamma}^t(0)$,
{\it i.e.}, the sum of Eqs.~(\ref{gagat}) and (\ref{eq:PiQ.2}),
with respect to the bare mass $m_t^0$.
When we express the differentiation with respect to $m_t^0$ in terms of
$m_t$, we pick up an additional finite contribution involving the
anomalous mass dimension $\gamma_m$,
\begin{equation}
\frac{m_t^0\partial}{\partial m_t^0} = \frac{1}{1+\gamma_m}\,
\frac{m_t\partial}{\partial m_t}.
\end{equation}
On the other hand, differentiation of $\Pi_{\gamma\gamma}^t(0)$  with respect
to $m_t$ yields
\begin{equation}
\label{betat}
\frac{m_t\partial}{\partial m_t} \Pi_{\gamma\gamma}^t(0)
= - \frac{\beta_{\alpha_{em}}^t}{\alpha_{em}},
\end{equation}
where $\beta_{\alpha_{em}}^t$ is the top--quark part of the QED $\beta$
function defined by $\mu \partial \alpha_{em}/\partial \mu =
\beta_{\alpha_{em}}$.
Combining these results with Eq.~(\ref{eq:L.HgagaLO}), we obtain the
effective $H\gamma\gamma$ interaction Lagrangian to all orders in $\alpha_s$
\cite{hggful,lhgamn,lhgamn2},
\begin{equation}
{\cal L}_{H\gamma\gamma} = F^{\mu\nu} F_{\mu\nu}\frac{H}{v}\,
\left( \frac{\beta_{\alpha_{em}}^t}{4\alpha_{em}}\,
\frac{1}{1+\gamma_m} - \frac{7\alpha_{em}}{8\pi} \right).
\label{eq:LHgan}
\end{equation}
The expansion of $\beta_{\alpha_{em}}^t$ up to ${\cal O}(\alpha_{em}\alpha_s)$
may be evaluated via Eq.~(\ref{betat}) and reads
\begin{equation}
\frac{\beta_{\alpha_{em}}^t}{\alpha_{em}} =N_c e_t^2\frac{2\alpha_{em}}{3\pi}
\left[1+\frac{3}{4}C_F\frac{\alpha_s}{\pi} + {\cal O}(\alpha_s^2) \right].
\end{equation}
Furthermore, the mass anomalous dimension of QCD is given by
\begin{equation}
\label{gammam}
\gamma_m = \frac{3}{2}C_F\frac{\alpha_s}{\pi} + {\cal O}(\alpha_s^2).
\end{equation}
Thus, in next--to--leading order, Eq.~(\ref{eq:LHgan}) becomes
\cite{hggful,lhgamn}
\begin{equation}
{\cal L}_{H\gamma\gamma} = \frac{\alpha_{em}}{2\pi}
F^{\mu\nu} F_{\mu\nu}\frac{H}{v}
\left[N_c\frac{e_t^2}{3} \left(1-\frac{3}{4}C_F\frac{\alpha_s}{\pi}\right)
-\frac{7}{4}\right].
\label{eq:L.Hgaga.2}
\end{equation}
This result is in agreement with the high--$m_t$ limit of the
two--loop QCD correction to the $H\gamma\gamma$ coupling
\cite{hggful,labj,hgaga}.

It is worthwhile to dwell on Eq.~(\ref{eq:LHgan}) for a little while.
This result can also be
obtained from a different type of LET, based on the trace anomaly of the
energy--momentum tensor, $\Theta^{\mu\nu}$.
It has been shown \cite{thetmymy} that the effective form of $\Theta^{\mu\nu}$
including all orders of the contributing couplings is given by
\begin{equation}
\label{thetamumu}
\Theta^\mu_\mu =  m_t^0 \bar t^0 t^0(1+\gamma_m) + \frac{\beta_{\alpha_{em}}}
{4\alpha_{em}} {\cal N} (F^{\mu\nu}F_{\mu\nu}),
\end{equation}
where ${\cal N} (\cdots)$ denotes the normal product.
Furthermore, it has been proven \cite{iwasaki} that the matrix element
$\langle 0|\Theta^\mu_\mu |\gamma\gamma \rangle$ vanishes at zero momentum
transfer,
\begin{equation}
\label{thegaga}
\lim_{Q^2\to 0} \langle 0|\Theta^\mu_\mu |\gamma\gamma \rangle = 0.
\end{equation}
Since a Higgs boson with vanishing momentum acts as a constant field,
it hence follows that
\begin{equation}
\label{themumuH}
\lim_{p_H\to 0} \langle H|\Theta^\mu_\mu H|\gamma\gamma \rangle = 0
\end{equation}
is fulfilled.
If we then multiply Eq.~(\ref{thetamumu}) by $H$ and substitute the result into
Eq.~(\ref{themumuH}), we obtain from
${\cal L}_{Ht\bar t}=-m_t^0 \bar t^0 t^0 H/v$ the effective $H\gamma\gamma$
Lagrangian as \cite{hggful,lhgamn,lhgamn2}
\begin{equation}
{\cal L}_{H\gamma\gamma} = F^{\mu\nu} F_{\mu\nu} \frac{H}{v}\,
\frac{\beta_{\alpha_{em}}^t}{4\alpha_{em}}\,\frac{1}{1+\gamma_m}.
\end{equation}
Here, we have exploited the facts that the operation in Eq.~(\ref{themumuH})
projects out the top--quark contribution to $\beta_{\alpha_{em}}$ and that,
in the low--energy limit, ${\cal N} (F^{\mu\nu}F_{\mu\nu})$ approaches the
corresponding operator containing the renormalized free fields.
This reproduces the first term of Eq.~(\ref{eq:LHgan}).
The $W$--boson contribution may be derived in a similar way.

{}From Lagrangian (\ref{eq:L.Hgaga.2}) we can read off the QCD corrections to
$\Gamma(H\to\gamma\gamma)$ and $\sigma(\gamma\gamma\to H)$, assuming that
$M_H\ll M_W,m_t$.
They are quite small, giving support to the notion that these processes are
theoretically well under control.

\subsubsection{Pseudoscalar Higgs Bosons}
%              =========================
According to the Adler--Bardeen theorem \cite{adlbar},
{\it i.e.}, the fact that the ABJ anomaly is not affected by renormalization,
the effective Lagrangian for the $A\gamma\gamma$ interaction is fixed
to all orders by Eq.~(\ref{eq:L.Agaga}).
Consequently, the two--loop QCD correction
to the $A\gamma\gamma$ coupling vanishes in the high--$m_t$ limit,
as may be also verified by explicit computation \cite{hggful,labj}.

\subsection{Higgs Couplings to Two Gluons}
%           =============================
\subsubsection{Scalar Higgs Bosons}
%              ===================
\paragraph{QCD Corrections.}
%          ================
In analogy to the top--quark--induced part of the effective $H\gamma\gamma$
Lagrangian~(\ref{eq:LHgan}), the QCD--corrected effective $Hgg$ Lagrangian of
the neutral scalar Higgs boson $H$ is given by \cite{hggful,lhgamn,lhgamn2}
\begin{equation}
\label{hggeff}
{\cal L}_{Hgg} =G^{a\mu\nu} G^a_{\mu\nu} \frac{H}{v}\,
\frac{\beta_{\alpha_s}^t}{4\alpha_s}\,\frac{1}{1+\gamma_m},
\end{equation}
where $\gamma_m$ is listed in Eq.~(\ref{gammam}) and
$\beta_{\alpha_s}^t$ denotes the top--quark contribution to the QCD
$\beta$ function defined by $\mu \partial \alpha_s /\partial \mu =
\beta_{\alpha_s}$.
Up to next--to--leading order, we have
\begin{equation}
\frac{\beta_{\alpha_s}^t}{\alpha_s} =  \frac{\alpha_s}{3\pi} \left[ 1 +
\frac{\alpha_s}{4\pi}(5N_c+3C_F) + {\cal O}(\alpha_s^2)\right].
\end{equation}
Thus, Eq.~(\ref{hggeff}) becomes \cite{hggful,lhgamn,lhgamn2}
\begin{equation}
\label{hggfinal}
{\cal L}_{Hgg} =\frac{\alpha_s}{12\pi} G^{a\mu\nu} G^a_{\mu\nu} \frac{H}{v}
\left[ 1 + \frac{\alpha_s}{4\pi}(5N_c-3C_F)+{\cal O}(\alpha_s^2) \right].
\end{equation}
This Lagrangian characterizes the $Hgg$ interaction in the theory
where the top quark has been integrated out.
For example, when we wish to compute the full two--loop QCD corrections to
$\Gamma(H\to gg)$, we just need to consider this Lagrangian in connection
with the usual Lagrangian of five--flavour QCD
and calculate the one--loop virtual correction and the tree--level
real correction.
The ultraviolet divergence of the virtual correction is removed by
renormalization, while the infrared and collinear singularities
cancel when the virtual and real corrections are combined.
The final result is \cite{hggful,lhgamn,inami}
\begin{equation}
\Gamma (H\to gg(g), gq\bar q) = \Gamma_{LO} (H\to gg) \left( 1+C_H
\frac{\alpha_s}{\pi} \right),
\label{eq:GHgg}
\end{equation}
where
\begin{eqnarray}
\Gamma_{LO}(H\to gg) & = & \frac{N_cC_F\alpha_s^2G_FM_H^3}{144\sqrt{2}\pi^3},
\\ \nonumber \\
\label{htogg}
C_H & = & \frac{103}{12}N_c - \frac{3}{2} C_F - \frac{7}{6} N_F
+ \frac{11N_c-2N_F}{6} \log \frac{\mu^2}{m_H^2},
\label{eq:Gam.ch}
\end{eqnarray}
with $N_F=5$ being the number of active quark flavours.
This agrees with the high--$m_t$ limit of the two--loop
calculation in six--flavour QCD \cite{hggful,lhgamn}.
The correction is quite sizeable; it increases the two--gluon decay rate of
the $\cal SM$ Higgs boson with intermediate mass by about 60\%.
This renders it more likely for this decay mode to
be detected at future $e^+e^-$ colliders.
If we keep the full mass dependence of $\Gamma_{LO}(H\to gg)$ \cite{glufus} in
Eq.~(\ref{eq:GHgg}), we obtain an approximation which, in the intermediate
Higgs--boson mass range, differs by at most 5\% from the exact result
\cite{hggful}.

Using Eq.~(\ref{hggfinal}), we can also calculate the QCD corrections to the
cross section of Higgs production via gluon fusion
\cite{hggful,lhgamn,lhgamn2,gghful}, which will be the primary source of
Higgs bosons at the LHC.
For the $\cal SM$ Higgs boson with intermediate mass,
these corrections range between 50\% and 80\%.
In this case, the high--$m_t$ limit provides a good approximation
\cite{hggful,gghful}.

\paragraph{Electroweak Corrections.}
%          ========================
By virtue of LET (\ref{eq:let.H}), we can also conveniently extract the
two--loop ${\cal O}(G_F m_t^2)$ electroweak correction to the $Hgg$ coupling.
Toward this end, we need to complement the one--loop top--quark
contribution to the gluon self--energy given in Eq.~(\ref{piggt}) with its
${\cal O}(G_F m_t^2)$ correction.
However, it turns out \cite{gghmt2} that the latter is ultraviolet finite,
provided that Eq.~(\ref{piggt}) is written with $m_t^0$,
and thus does not contribute upon application of LET~(\ref{eq:let.H}).
Therefore, we just need to renormalize the top--quark mass appearing in the
one--loop calculation to ${\cal O}(G_Fm_t^2)$.
In the on--shell scheme, this may be achieved by substituting
\begin{equation}
\label{mshift}
m_t^0  =  m_t + \delta m_t,
\end{equation}
with \cite{hbb1}
\begin{equation}
\frac{\delta m_t}{m_t}  =  x_t
\left(\frac{4\pi\mu^2}{m_t^2} \right)^{\epsilon}\Gamma (1+\epsilon)
\left[ \frac{3}{2\epsilon} + 4 + {\cal O}(\epsilon) \right],
\end{equation}
where $x_t$ is defined below Eq.~(\ref{deltahbb}).
In this way, we find the non--universal piece,
\begin{equation}
\delta_{Hgg,\,nu} = -3 x_t.
\end{equation}
Combining this with the universal part $\delta_u$ given in Eq.~(\ref{deltau}),
we obtain the ${\cal O}(G_F m_t^2)$ term to be included within the square
brackets of Eq.~(\ref{hggfinal}),
\begin{equation}
\delta_{Hgg,\,ew}= \delta_{Hgg,\,nu}+\delta_u
=\left(\frac{7}{6}N_c-3\right)x_t.
\end{equation}
This is in agreement with Ref.~\cite{gghmt2}.
The corresponding correction factor for $\Gamma(H\to gg)$ and $\sigma(gg\to H)$
is then $(1+2\delta_{Hgg,\,ew})$, which leads to an insignificant increase, by
about three tenths of a percent.

\subsubsection{Pseudoscalar Higgs Bosons}
%              =========================
In Section~4.1.2, we have seen that, as a consequence of the
Adler--Bardeen theorem \cite{adlbar}, the $A\gamma\gamma$ coupling does not
receive QCD corrections in the high--$m_t$ limit.
This also holds true for the $Agg$ interaction.
Thus, the two--loop QCD correction to $\Gamma(A\to gg)$ may be computed in
five--flavour QCD by using the effective Lagrangian (\ref{eq:L.Agg}).
The result is \cite{hggful,labj}
\begin{equation}
\Gamma(A\to gg(g), gq\bar q) = \Gamma_{LO} (A\to gg) \left( 1+C_A
\frac{\alpha_s}{\pi} \right),
\end{equation}
where
\begin{eqnarray}
\Gamma_{LO} (A\to gg) & = & \frac{N_cC_F\alpha_s^2G_FM_A^3}{64\sqrt{2}\pi^3},
\\ \nonumber \\
\label{atogg}
C_A & = & \frac{97}{12}N_c -\frac{7}{6} N_F + \frac{11N_c-2N_F}{6} \log
\frac{\mu^2}{m_A^2},
\end{eqnarray}
with $N_F=5$.
For an intermediate--mass $A$ boson, this correction amounts to about 60\%.
One should bear in mind that, in the ${\cal MSSM}$, this result is only
reliable for small values of $\tan\beta$, of order unity, where the top--quark
contribution is dominant.

The QCD correction to $\sigma(gg\to A)$ in the high--$m_t$ limit may be
computed in a similar way.
The $gg\to A$ mechanism will be the dominant source of $A$ bosons at the LHC
\cite{hgamlhc}.
The QCD correction turns out to be 50--100\% \cite{hggful,gga2}.

\subsection{Higgs Couplings to One $Z$ Boson and One Photon}
%           ===============================================
\subsubsection{Scalar Higgs Bosons}
%              ===================
In Section~4.1.1, we have derived the two--loop QCD correction to the
top--quark--induced part of the $H\gamma\gamma$ coupling by applying
LET~(\ref{eq:let.H}) to the respective contribution to the photon
self--energy.
The corresponding result for the $HZ\gamma$ interaction follows by simply
adjusting the coupling constants.
The resulting QCD correction may be accommodated in Eq.~(\ref{eq:L.HZgaLO})
by multiplying the first term contained within the parentheses with the
factor $[1-3C_F\alpha_s/(4\pi)]$.
This agrees with the leading high--$m_t$ term of the full two--loop calculation
\cite{hzga2}.

\subsubsection{Pseudoscalar Higgs Bosons}
%              =========================
{}From arguments similar to those in Sections~4.1.2 and 4.2.2 it follows on
that
the effective $AZ\gamma$ Lagrangian does not receive any QCD corrections
in the high--$m_t$ limit.
At the two--loop order, this has been checked by an explicit analysis.

\subsection{$H\to b \bar b$}
%           ===============
The ${\cal O}(G_Fm_t^2)$ analysis of Section~3.4 readily carries over to
${\cal O}(\alpha_sG_Fm_t^2)$.
We just need to evaluate the ${\cal O}(\alpha_sG_Fm_t^2)$ corrections to
Eqs.~(\ref{bsigs}) and (\ref{bsigv}).
These read \cite{hbb2}
\begin{eqnarray}
m_b^0 \Sigma_S^{(2)} (0) & = & C_F \frac{\alpha_s}{\pi}\,
\frac{g_t^0}{(4\pi)^2} \left[
\frac{4\pi\mu^2}{(m_t^0)^2}\right]^{2\epsilon}\Gamma^2(1+\epsilon)
\left[ -\frac{3}{2\epsilon^2}g_b^0 m_t^0
+\frac{1}{\epsilon} \left(  \frac{3}{4} g_t^0 m_b^0 - 2g_b^0m_t^0 \right)
\right. \nonumber \\
& & \hspace{6cm} + \left.\vphantom{\frac{3}{2\epsilon^2}}
{\cal O} (1) \right], \nonumber \\ \nonumber \\
\Sigma_V^{(2)} (0) & = & C_F \frac{\alpha_s}{\pi}\,
\frac{(g_t^0)^2}{(4\pi)^2} \left[ \frac{4\pi
\mu^2}{(m_t^0)^2}\right]^{2\epsilon}\Gamma^2(1+\epsilon)
 \left[ \frac{3}{8\epsilon^2} +
\frac{1}{8\epsilon} + {\cal O} (1) \right].
\label{eq:bself2}
\end{eqnarray}
{}From the first line of Eq.~(\ref{delhbbzero}) we then
obtain the ${\cal O}(\alpha_sG_Fm_t^2)$ term of $\delta_{Hb\bar b}^0$ in
Eq.~(\ref{lhbb}),
\begin{equation}
\delta_{Hb\bar b}^{0(2)}  =   -\Sigma_S^{(2)}(0)
+4\epsilon\left[\Sigma_S^{(2)}(0)+\Sigma_V^{(2)}(0)\right].
\end{equation}
Again, the $\Sigma_S (0)$ term is removed by the bottom--quark mass and
wave--function renormalizations of Eq.~(\ref{brenor}).
Since we are now working at next--to--leading order, we also need to
renormalize the top--quark mass in the leading--order expressions, {\it i.e.},
we need to use Eq.~(\ref{mshift}) in connection with
\begin{equation}
\frac{\delta m_t}{m_t}  =  -C_F \frac{\alpha_s}{\pi}
\left( \frac{4\pi\mu^2}{m_t^2}\right)^\epsilon \Gamma(1+\epsilon) \left[
\frac{3}{4\epsilon} + 1 + 2\epsilon + {\cal O}(\epsilon^2)\right].
\label{eq:mtCT}
\end{equation}
This renders the non--universal ${\cal O}(\alpha_sG_Fm_t^2)$ correction
finite \cite{hbb2},
\begin{equation}
\delta_{Hb\bar b,\,nu}^{(2)} = \frac{3}{4} C_F \frac{\alpha_s}{\pi}x_t.
\end{equation}
We still need to renormalize the wave function and vacuum expectation value
of the Higgs field in ${\cal O}(\alpha_sG_Fm_t^2)$.
This yields \cite{hll2,karlsruhe}
\begin{equation}
\label{deltautwo}
\delta_u^{(2)} = -\frac{1}{2}\left[\zeta(2)+\frac{3}{2}\right]
N_cC_F\frac{\alpha_s}{\pi}x_t.
\end{equation}
Thus, the ${\cal O}(\alpha_sG_Fm_t^2)$ term to be included in
Eq.~(\ref{hbbeff}) comes out as \cite{hbb2}
\begin{equation}
\delta_{Hb\bar b}^{(2)} =  \delta_{Hb\bar b,\,nu}^{(2)} + \delta_u^{(2)}
=\left\{\frac{3}{4}-\frac{1}{2}\left[\zeta(2)+\frac{3}{2}\right]N_c\right\}
C_F\frac{\alpha_s}{\pi}x_t.
\end{equation}

The ${\cal O}(\alpha_sG_Fm_t^2)$ correction to $\Gamma(H\to b\bar b)$ receives
an additional contribution from the interference of the ${\cal O}(G_Fm_t^2)$
term (\ref{delhbb}) and the well--known ${\cal O}(\alpha_s)$ correction
\cite{hbbqcd}.
In the limit $m_b\ll M_H$, the latter is given by \cite{hbbqcd}
\begin{equation}
\delta_{QCD} = \frac{3}{2} C_F \frac{\alpha_s}{\pi} \left( \frac{3}{2}
- \log \frac{M_H^2}{m_b^2} \right),
\end{equation}
where it is understood that the Born formula for $\Gamma(H\to b\bar b)$ is
written with the bottom--quark pole mass $m_b$.
The combined correction is \cite{hbb2,karlsruhe}
$(1+\delta_{Hb\bar b})^2(1+\delta_{QCD})$.
Numerically, $\delta_{Hb\bar b}^{(2)}$ reduces the effect of
$\delta_{Hb\bar b}^{(1)}$ by about 40\%, so that the $m_t$ dependence of
$\Gamma(H\to b\bar b)$ is weakened significantly.
It is well known that the large logarithm of $\delta_{QCD}$ may be absorbed
into the running bottom--quark mass evaluated at scale $M_H$ \cite{hbbqcd}.

\subsection{$HZZ$ coupling and $e^+e^- \to ZH$}
%           =============================
Also the ${\cal O}(G_Fm_t^2)$ analysis of Section~3.5 may be
straightforwardly extended to ${\cal O}(\alpha_sG_F$      $m_t^2)$.
This only requires knowledge of the ${\cal O}(\alpha_sG_Fm_t^2)$ term
of $\Pi_{ZZ}(0)$, which may be extracted from Ref.~\cite{drho2},
\begin{equation}
\Pi^{(2)}_{ZZ} (0) = N_c C_F \frac{\alpha_s}{\pi} (g_Z^0v^0)^2 x_t^0 \left[
\frac{4\pi\mu^2}{(m_t^0)^2} \right]^{2\epsilon} \Gamma^2(1+\epsilon) \left[
-\frac{3}{2\epsilon^2} + \frac{7}{4\epsilon} - \frac{1}{8} +{\cal O}
(\epsilon) \right].
\end{equation}
Proceeding as in the one--loop case and renormalizing the top--quark mass
in the ${\cal O}(G_Fm_t^2)$ terms according to Eq.~(\ref{eq:mtCT}),
we find the non--universal correction to be \cite{lhzz2}
\begin{equation}
\delta_{HZZ,\,nu}^{(2)} = \frac{9}{2} N_c C_F \frac{\alpha_s}{\pi} x_t.
\end{equation}
Combining this with Eq.~(\ref{deltautwo}), we get
\begin{equation}
\label{deltahzztwo}
\delta_{HZZ}^{(2)} = \delta_{HZZ,\,nu}^{(2)}+\delta_u^{(2)}=
\frac{1}{2}\left[ \frac{15}{2} - \zeta (2) \right] N_c C_F
\frac{\alpha_s}{\pi}x_t,
\end{equation}
which upgrades the effective $HZZ$ Lagrangian (\ref{eq:LHZZ1}) to
${\cal O}(\alpha_sG_Fm_t^2)$.

Using Eq.~(\ref{deltahzztwo}) together with the well--known
${\cal O}(\alpha_sG_Fm_t^2)$ correction to the $\rho$ parameter \cite{drho2},
\begin{equation}
\Delta \rho^{(2)} = - \left[\zeta (2)+\frac{1}{2} \right]
N_c C_F \frac{\alpha_s}{\pi} x_t,
\end{equation}
we obtain from Eq.~(\ref{deltahzee})
the corresponding correction to $\sigma (e^+ e^-\to ZH)$ \cite{lhzz2},
\begin{equation}
\delta_{HZe^+ e^-}^{(2)} =  N_c C_F \frac{\alpha_s}{\pi} x_t
\left\{ 7 - 2 \zeta (2)
+ 8 \frac{c_w^2 Q_e v_e}{v_e^2 + a_e^2}
 \left[ \zeta (2) + \frac{1}{2} \right]\right\}.
\end{equation}
The QCD correction screens the leading ${\cal O}(G_F m_t^2)$ term by
about 20\% and thus reduces the sensitivity to the top--quark mass.

\subsection{$HWW$ coupling and $e^+e^- \to \nu_e\bar\nu_eH$}
%           ==========================================
The extension of the effective $HWW$ Lagrangian (\ref{eq:LHWW1}) to
${\cal O}(\alpha_sG_Fm_t^2)$ proceeds quite similarly to the $HZZ$ case, and
we merely list the starting point and the final result.
The ${\cal O}(\alpha_sG_Fm_t^2)$ contribution to $\Pi_{WW}(0)$ may be found
in Ref.~\cite{drho2},
\begin{equation}
\Pi^{(2)}_{WW} (0) = N_c C_F \frac{\alpha_s}{\pi} (g_W^0v^0)^2 x_t^0 \left[
\frac{4\pi\mu^2}{(m_t^0)^2} \right]^{2\epsilon} \Gamma^2(1+\epsilon) \left[
-\frac{3}{2\epsilon^2} + \frac{1}{4\epsilon} + \zeta (2) - \frac{7}{8}
+{\cal O}(\epsilon)\right].
\end{equation}
The final result is new and reads
\begin{equation}
\delta_{HWW}^{(2)} =  \frac{1}{2}
\left[ \frac{9}{2} - \zeta (2) \right] N_c C_F \frac{\alpha_s}{\pi}x_t.
\end{equation}
This reduces the magnitude of the ${\cal O}(G_F m_t^2)$ term of
Eq.~(\ref{deltahww}) by about 8\%.
As we have seen in Section~3.6,
in the on--shell scheme formulated with $G_F$,
$\sigma(e^+e^-\to H\nu_e\bar\nu_e)$ is corrected
by the factor $(1+2\delta_{HWW})$.

\section{Conclusions}
%        ===========
In this paper, we have reviewed low--energy theorems for the
evaluation of one--loop amplitudes with light Higgs bosons as external
particles.
We have then shown how these theorems may be extended to the two--loop order.
These theorems allow us to construct effective Lagrangians for the
interactions of the Higgs bosons with other light particles by integrating out
the heavy loop particles.
We have demonstrated the usefulness of these theorems for practical
calculations by presenting a variety of applications to Higgs--boson production
and decay processes which will be of major phenomenological relevance at future
colliding--beam experiments.

\vspace{1cm}

\noindent
{\bf Acknowledgements.} We are grateful to W.A.~Bardeen, A.~Djouadi,
B.~Grz\c adkowski, W.~Kilian, and P.M.~Zerwas for illuminating discussions.
We owe special thanks to A.~Djouadi for providing us with details
of Ref.~\cite{gghmt2}. We thank W.A.~Bardeen, A.~Djouadi, and P.M.~Zerwas for
critically reading the manuscript.
One of us (BAK) is indebted to the FNAL Theory Group for inviting him as a
Guest Scientist and for the great hospitality extended to him.

\end{document}